\documentclass[aps,prl,floatfix,nofootinbib,showpacs,twocolumn]{revtex4}
\usepackage{graphicx}
\usepackage{amsmath}
\usepackage{amssymb}

\renewcommand{\u}{\bar{u}}
\renewcommand{\v}{\bar{v}}

\newcommand{\BEQ}{\begin{eqnarray}}
\newcommand{\EEQ}{\end{eqnarray}}
\newcommand{\BEA}{\begin{eqnarray}}
\newcommand{\EEA}{\end{eqnarray}}

\renewcommand{\d}{{\rm d}}

\newcommand{\eps}{\varepsilon}

\newcommand{\tr}{{\rm tr}}

\renewcommand{\S}{{\rm\bf C}}
\newcommand{\R}{{\rm\bf H}}

\newcommand{\diag}{{\rm diag}}
\newcommand{\ssum}{{\sum}}
\newcommand{\comment}[1]{}
\begin{document} 
\draft
\title
{Optimal thermal refrigerator.}
\date{\today}
\author{ Armen E. Allahverdyan$^1$, Karen Hovhannisyan$^1$, Guenter Mahler$^2$}
\affiliation{$^1$Yerevan Physics Institute,
Alikhanian Brothers Street 2, Yerevan 375036, Armenia,\\
$^2$Institute of Theoretical Physics I, University of Stuttgart, 
Pfaffenwaldring 57, 70550 Stuttgart, Germany}

\begin{abstract} We study a refrigerator model which consists of two
$n$-level systems interacting via a pulsed external field. Each system
couples to its own thermal bath at temperatures $T_h$ and $T_c$,
respectively ($\theta\equiv T_c/T_h<1$). The refrigerator functions in
two steps: thermally isolated interaction between the systems driven by
the external field and isothermal relaxation back to equilibrium.  There
is a complementarity between the power of heat transfer from the cold
bath and the efficiency: the latter nullifies when the former is
maximized and {\it vice versa}. A reasonable compromise is achieved by
optimizing over the inter-system interaction and intra-system energy
levels the product of the heat-power and efficiency. The efficiency is
then found to be bounded from below by $\zeta_{\rm
CA}=\frac{1}{\sqrt{1-\theta}}-1$ (an analogue of Curzon-Ahlborn
efficiency for refrigerators), besides being bound from above by the
Carnot efficiency $\zeta_{\rm C} = \frac{1}{1-\theta}-1$. The lower
bound is reached in the equilibrium limit $\theta\to 1$, while the
Carnot bound is reached (for a finite power and a finite amount of
heat transferred per cycle) in the macroscopic limit $\ln n\gg 1$. The
efficiency is exactly equal to $\zeta_{\rm CA}$, when the above
optimization is constrained by assuming homogeneous energy spectra for both
systems.

\end{abstract}


\comment{
Energy conversion, 84.60.-h

Irreversible thermodynamics, 05.70.Ln

Thermodynamics, 05.70.-a

Heat engines, 07.20.Pe

Refrigeration, 07.20.Mc

quantum statistical mechanics, 05.30.-d

}

\maketitle

Thermodynamics studies principal limitations imposed on the performance
of thermal machines, be they macroscopic heat engines or refrigerators
\cite{callen}, or small devices in nanophysics \cite{q_t} and biology
\cite{venturi}. Let us recall three basic definitions applicable to any
thermal machine taking as an example a refigerator driven by a source of
work: {\it i)} heat $Q_c$ transferred per cycle of operation from a cold
body at temperature $T_c$ to a hot body at temperature $T_h$
($T_h>T_c$).  {\it ii)} Power, which is $Q_c$ divided over the cycle
duration $\tau$.  {\it iii)} Efficiency (or performance coefficient)
$\zeta=Q_c/W$, which quantifies the useful output $Q_c$ over the work
$W$ spent by the work-source for making the cycle. The second law
imposes the Carnot bound $\zeta\leq \zeta_{\rm C} = T_c/(T_h-T_c)$ on
the efficiency of refrigeration \cite{callen}. Within the usual
thermodynamics the Carnot bound is reached only for a reversible, i.e.,
an infinitely slow process, which means it is reached at zero power
\cite{callen}. The practical value of the Carnot bound is frequently
questioned on this ground. The drawback of zero power is partially cured
within finite-time thermodynamics (FTT), which is still
based on the quasi-equilibrium concepts \cite{ftt}. For heat-engines FTT gives
an upper bound $\eta\leq \eta_{\rm
CA}\equiv 1-\sqrt{T_c/T_h}$ (Curzon-Ahlborn, or CA efficiency) for the
efficiency $\eta$ at the maximal power of work-extraction
\cite{ca,broek}. Naturally, $\eta_{\rm CA}$ is smaller than the Carnot
upper bound $1-T_c/T_h$ for heat-engines. 

Heat engines have recently been studied within microscopic theories,
where one is easily able to go beyond the quasi-equilibrium regime
\cite{kosloff,armen,tu,izumida_okuda,udo,esposito,jmod,henrich}. For
certain classes of heat-engines the CA efficiency is a {\it lower} bound
for the efficiency at the maximal power of work
\cite{armen,tu,izumida_okuda}. This bound is reached at the
quasi-equilibrium situation $T_h\to T_c$ in agreement with the finding of
FTT. The result is consistent with other studies \cite{udo,esposito}. 

The situation with refrigerators at a finite power is less clear, though
\cite{yan_chen,velasco,jimenez,unified}. Here maximizing the power of cooling
does not lead to reasonable results, since there is an additional
complementarity (not present for heat engines) \cite{velasco}: when
maximizing the power one simultaneously minimizes the efficiency to
zero, and {\it vice versa}. 

We study optimal regimes of finite-power refrigeration via a
realistic model, which can be optimized over almost all of its
parameters. The model is quantum, but it admits a
classical interpretation. The interest in small-scale refrigerators is
triggered by the importance of of cooling processes for functioning of small devices
and for displaying quantum features of matter
\cite{q_t,henrich,feldman,segal,rezek}. 

\comment{ After introducing the model
and confirming the heat-power-efficiency complementarity, we show that
within the present apporach the most meaningful way of optimizing its
functioning is to maximize the product of efficiency and the heat power.
This leads to a lower bound $\zeta_{ref}= -1+1/\sqrt{1-\theta}$ for the efficiency,
in addition to the upper Carnot bound $\zeta = \frac{\theta}{1-\theta}$. The
same expression $\zeta_{ref}$ was obtained within FTT as an
upper bound, when maximizing the ratio of efficiency and the cycle time
\cite{velasco}. }

Consider two quantum systems $\R$ and $\S$ with Hamiltonians $H_\R$
and $H_\S$, respectively. Each system has $n$ energy levels. 
Initially, $\R$ and $\S$ do not interact and are in equilibrium
at temperatures $T_h=1/\beta_h>T_c=1/\beta_c$:
\BEA
\label{1}
\rho={e^{-\beta_h H_\R}}/{\tr\, [e^{-\beta_h H_\R}]}, ~~
\sigma={e^{-\beta_c H_\S}}/{\tr\, [e^{-\beta_c H_\S}]}, 
\EEA
where $\rho$ and $\sigma$ are the initial Gibbsian density matrices of $\R$ and $\S$, respectively. 
We write
\BEA
\rho=\diag [r_n,...,r_1], \quad r_n\leq ...\leq r_1, \nonumber\\
\label{2}
\sigma=\diag [s_n,...,s_1], \quad s_n\leq ...\leq s_1,\\
H_\R=\diag [\eps_n,...,\eps_1=0\,],\,\, \eps_n\geq ...\geq \eps_1, \nonumber\\
H_\S=\diag [\mu_n,...,\mu_1=0\,], \,\, \mu_n\geq ...\geq \mu_1,
\label{5}
\EEA
where $\diag[a,..,b]$ is a diagonal matrix with entries $(a,...,b)$, and
where without loss of generality we have nullified the lowest energy level
of both $\R$ and $\S$.  Thus the overall initial density matrix is 
$\Omega_{\rm in}=\rho\otimes\sigma$, and the initial Hamiltonian 
$H_\R\otimes 1+1\otimes H_\S $. 

\begin{figure}[ht]
\includegraphics[width=7.5cm]{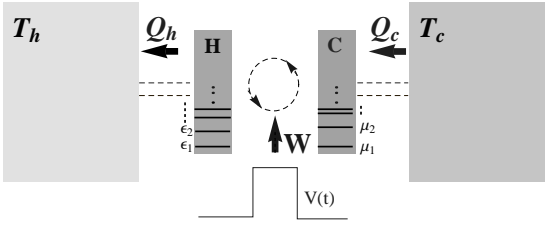}
\caption{ The refrigerator model. Two systems $\R$ 
and $\S$ operate between two baths at temperatures
$T_c<T_h$ and are driven by an external potential $V(t)$. 
$W$ and $Q_c$ and $Q_h$ are, respectively, the work put into
the overall system and the heats transferred from the cold bath and to the hot bath. } 
\label{f1}
\end{figure}

The goal of any refrigerator is to transfer heat from the cooler bath to
the hotter one at the expense of consuming work from an external source.
The present refrigerator model functions in the following two steps; 
see Fig.~\ref{f1}. 

{\bf 1.} $\R$ and $\S$ interact with each other and with external
sources of work. The overall interaction is described via a
time-dependent potential $V(t)$ in the total Hamiltonian $H_\R\otimes
1+1\otimes H_\S+V(t)$ of $\R+\S$. The interaction process is thermall
isolated: $V(t)$ is non-zero only in a short time-window $0\leq t\leq
\delta$ and is so large there that the influence of all other couplings
[e.g., couplings to the baths] can be neglected [pulsed regime]. Thus
the dynamics of $\R+\S$ is unitary for $0\leq t\leq \delta$:
\BEA
\Omega_{\rm f}\equiv \Omega (\delta )={\cal U}\,
\Omega_{\rm i}\, {\cal U}^\dagger, \qquad  {\cal U}={\cal T}e^{-\frac{i}{\hbar} \int_0^\delta \d s V(s)  }
\label{uni}
\EEA
where $\Omega_{\rm i}=\Omega(0)=\rho\otimes\sigma$ is the initial state defined in
(\ref{1}), $\Omega_{\rm f}$ is the final density matrix, ${\cal
U}$ is the unitary evolution operator, and where ${\cal T}$ is the time-ordering operator. 
The work put into $\R+\S$ in this process is \cite{callen}
\BEA
W=E_{\rm f}-E_{\rm i}=\tr [\,( H_\R\otimes 1+1\otimes H_\S)\, (\Omega_{\rm f}-\Omega_{\rm i}) \,],
\label{work}
\EEA
where $E_{\rm f}$ and $E_{\rm i}$ are initial and final energies of $\R+\S$.

{\bf 2.} Once the overall system $\R+\S$ arrives at the final state
$\Omega_{\rm fin}$, $V(t)$ is switched off, and $\R$ and $\S$ (within
some relaxation time) return back to their initial states (\ref{1})
under influence of the hot and cold thermal baths, respectively.  Thus
the cycle is complete and can be repeated again.  Because the energy is
conserved during the relaxation, the hot bath gets an amount of heat
$Q_h$, while the cold bath gives up the amount of heat $Q_c$
\BEA
\label{heats}
Q_h=\tr (H_\R[\,{\rm tr}_\S\Omega_{\rm f}-\rho]),\, \, \,\,
Q_c=\tr (H_\S[\sigma-{\rm tr}_\R\Omega_{\rm f}]),
\EEA
where ${\rm tr}_\R$ and ${\rm tr}_\S$ are the partial traces.
Eq.~(\ref{1}) and the unitarity (\ref{uni}) lead to
\BEA 
\label{43}
\beta_h Q_h-\beta_c Q_c=S(\Omega_{\rm f}||\Omega_{\rm i})\equiv {\rm tr}[\Omega_{\rm f}\ln \Omega_{\rm f}
-\Omega_{\rm f}\ln \Omega_{\rm i}],
\EEA 
where $S(\Omega_{\rm f}||\Omega_{\rm i})\geq 0$ is the relative entropy. This quantity nullifies if and only if 
$\Omega_{\rm f}=\Omega_{\rm i}$; otherwise it is positive. Eq.~(\ref{43}) is the Clausius inequality,
with $S(\Omega_{\rm f}||\Omega_{\rm i})\geq 0$ quantifying the entropy production.
Eqs.~(\ref{work}--\ref{43}) and the energy conservation $Q_h=W+Q_c$ imply
$(\beta_c-\beta_h)Q_c\leq\beta_h W$, meaning that
in the refrigeration regime we have $Q_c>0$ and thus $W>0$.
Eq.~(\ref{43}) leads to the Carnot bound for the
efficiency $\zeta$ of our refrigerator
\BEA
\zeta\,\equiv \,{Q_c}/{W}\,
\,\leq\,\,{\theta}/{(1-\theta)}\,\equiv\,\zeta_{\rm C}, ~~~ \theta\,\equiv\, T_c/T_h<1.~
\EEA
Recall that the power of refrigeration is defined as the
ratio of the transferred heat to the cycle duration $\tau$, $Q_c/\tau$. For the
present model $\tau$ is mainly the duration of the second stage, i.e.,
$\tau$ is the relaxation time, which depends on the concrete physics of the
system-bath coupling. For a weak system-bath coupling $\tau$ is larger
than the internal characteristic time of $\R$ and $\S$. In contrast, for the
collisional system-bath interaction, $\tau$ can be very short; see,
e.g., \cite{armen} for a detailed discussion. Thus in
our setup the cycle time $\tau$ is finite, and the power of
refrigeration $Q_c/\tau$ does not vanish due to a large cycle time,
though it can vanish due to $Q_c\to 0$. 

We now proceed to optimize the functioning of the refrigerator over the
three sets of available parameters: the energy spacings
$\{\eps_k\}_{k=2}^n$, $\{\mu_k\}_{k=2}^n$, and the unitary operators
(\ref{uni}) [or the interaction Hamiltonian $V(t)$].
  
We start by maximizing the transferred heat $Q_c=\tr
(H_\S[\sigma-{\rm tr}_\R\Omega_{\rm f}])$, which is the main characteristics of the
refrigerator. 
Note that the initial energy $\tr [H_\S\sigma]$ depends only on
$\{\eps_k\}_{k=2}^n$. Therefore, we first choose $\{\mu_k\}_{k=2}^n$ and
$V(t)$ so that the final energy $\tr [H_\S\Omega_{\rm f}]$ attains its
minimal value equal to zero. 
Then we 
maximize $\tr [H_\S\sigma]$ over $\{\eps_k\}_{k=2}^n$. Note from (\ref{2}, \ref{5})
\BEA
1\otimes H_\S &=& {\rm diag}[\,\mu_1\,\,,\ldots,\,\,\,\mu_1,\ldots,
                             \,\,\mu_n\,\,\,,\ldots,\,\,\mu_n\,\, ],\nonumber\\
\Omega_{\rm i}=\rho\otimes\sigma &=& {\rm
  diag}[\, s_1r_1,\ldots,s_1r_n,\ldots,
                            s_nr_1,\ldots,s_nr_n\, ].\nonumber
\EEA
It is clear that $\tr [H_\S\Omega_{\rm f}]=\tr [H_\S{\cal U}\Omega_{\rm
 i}{\cal U}^\dagger]$ goes to zero when, e.g., $r_2=\ldots=r_n\to 0$ 
($\eps\equiv\eps_2=\ldots=\eps_n\to \infty$),
while ${\cal U}$ amounts to the SWAP operation 
${\cal U}\rho\otimes\sigma{\cal U}^\dagger = \sigma\otimes\rho$. 
It is checked by a direct
inspection that the maximization of the initial energy 
$\tr [H_\S\sigma]$ over $\{\eps_k\}_{k=2}^n$ produces the same
structure of $n-1$ times degenerate upper energy levels
$\mu\equiv\mu_2=\ldots=\mu_n$. Denoting 
\BEA
\label{burundi}
v\equiv s_2=..=s_n= e^{-\beta_c \mu},~~ u\equiv r_2=..=r_n= e^{-\beta_h \eps},
\EEA
we obtain for $Q_c$
\BEA
\label{10} 
Q_c=
T_c \ln\left[\frac{1}{v}\right]\,\frac{(v-u)(n-1) }{[\,1+(n-1)v\,][\,1+(n-1)u\,]}, 
\EEA
where according to the above discussion, $Q_c$ is maximized for $u\to
0$, and where $v$ is to be found from maximizing $Q_c|_{u\to 0}$ in 
(\ref{10}) over $v$, i.e., $v$ is
determined via $1+(n-1)v+\ln v=0$. Thus $\S$ can be cooled down to its ground state, but 
at a vanishing efficiency.

For the efficiency we get for the present situation ($\R$ and $\S$
have $n-1$ times degenerate upper levels, while ${\cal U}$ amounts to
the SWAP operation):
\BEA \label{11}
\zeta\,=\,{Q_c}/{W}\,
=\,\theta\, \ln[\,v\,]\, \left(\,\ln[\,u\,]-\theta\ln[\,v\,]\,\right)^{-1}.
\label{peshavar}
\EEA
The maximization of $Q_c$ leads to $u\to 0$, which then
means that $\zeta$ in (\ref{11}) goes to zero.
Note that $\zeta$ in (\ref{11}) reaches its maximal Carnot value
$\theta/(1-\theta)$ for $u=v$, which nullifies the transferred heat
$Q_c$; see (\ref{10}). Now we show
that $Q_c$ tends to zero
upon maximizing $\zeta$ over {\it all} free parameters
$\{\eps_k\}_{k=2}^n$, $\{\mu_k\}_{k=2}^n$ and ${\cal U}$.
Denoting $\{| i_\R\rangle \}_{k=1}^n$ and $\{| i_\S\rangle \}_{k=1}^n$
for the eigenvectors of $H_\R$ and $H_\S$, respectively, we note 
from (\ref{work}, \ref{heats}) 
that $W$ and $Q_c$ feel ${\cal U}$ only via 
$C_{ij\,|\,kl}=|\langle i_\R j_\S |{\cal U}| k_\R l_\S\rangle|^2  $.
This matrix is double-stochastic \cite{olkin}: 
${\ssum}_{ij }C_{ij\,|\,kl}={\ssum}_{kl }C_{ij\,|\,kl}=1$.
Conversely, for {\it any} double-stochastic matrix $C_{ij\,|\,kl}$ there
is {\it some} unitary matrix $U$ with matrix elements $U_{ij\,|\,kl}$,
so that $C_{ij\,|\,kl}=|U_{ij\,|\,kl}|^2$ \cite{olkin}. Thus, when
maximizing various functions of $W$ and $Q_c$ over the unitary ${\cal
U}$, we can directly maximize over the $(n^2-1)^2$ independent elements
of $n^2\times n^2$ double stochastic matrix $C_{ij\,|\,kl}$.  

We did not find an analytic way of carrying out the complete
maximization of $\zeta$ over all free parameters. Thus we had to rely on
numerical recipes of Mathematica 7, which for $n=1,\ldots, 5$ confirmed
that $Q_c$ nullifies whenever $\zeta$ reaches (along any path) its
maximal Carnot value. We believe this holds for an arbitrary $n$, though
we lack any rigorous prove of this assertion.

\begin{figure}[ht]
\includegraphics[width=6cm]{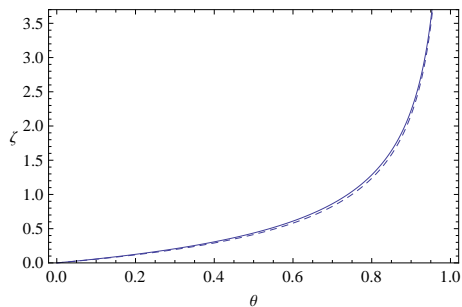}
\caption{ Solid line: efficiency $\zeta$ of the optimized refrigerator versus the temperature ratio $\theta=
T_c/T_h$ for $n=3$; see (\ref{11}). In the scale of this figure $\zeta(n=2)$ and $\zeta(n=3)$ are almost indistinguishable.
Dashed line: the lower bound $\frac{1}{\sqrt{1-\theta}}-1$. } 
\label{f2}
\end{figure}

Thus, neither $Q_c$ nor $\zeta$ are good target quantities for determining
an optimal regime of refrigeration. But $\chi \equiv Q_c\zeta$
is such a target quantity, as will be seen shortly. This is the most natural
choice for our setup. This choice was also employed in \cite{yan_chen}. Refs.~\cite{unified,feldman} 
report on other approaches to defining the optimal refrigeration.

The numerical maximization of $\chi=\zeta Q_c$ over
$\{\eps_k\}_{k=2}^n$, $\{\mu_k\}_{k=2}^n$ and ${\cal U}$ has been
carried out for $n=1,\ldots,5$ along the above lines. It produced the
same structure: both $\R$ and $\S$ have $n-1$ times degenerate upper
levels, see (\ref{burundi}), and the optimal ${\cal U}$ again
corresponds to SWAP operation. We thus get for $\chi=\zeta Q_c$ [see
(\ref{10}, \ref{11})]
\BEA
\label{14}
\chi(\u,\v)= 
\frac{T_c\theta (n-1)(\v-\u)\ln^2\frac{1}{\v}}
{[\ln\frac{1}{\u}-\theta\ln\frac{1}{\v}][1+(n-1)\u][1+(n-1)\v]},
\EEA
where $\u$ and $\v$ are found from maximizing $\chi(u,v)$ via
$\partial_u \chi=\partial_v \chi=0$. Though we have numerically checked these
results for $n\leq 5$ only, we again trust that they hold for an
arbitrary $n$ (one can, of course, always consider the above structure
of energy spacings and ${\cal U}$ as a useful ansatz).  Note that $\u$
and $\v$ depend on $\theta=T_c/T_h$. The efficiency $\zeta$ and the
transferred heat $Q_c$ are given by (\ref{11}) and (\ref{10}) with
$u\to\u$ and $v\to\v$; see Fig.~\ref{f2}.

Since the
state of $\R+\S$ after the action of $V(t)$ is $\sigma\otimes\rho$, and
because in the optimal regime the upper level for both $\R$ and $\S$ is
$n-1$ times degenerate, one can introduce non-equilibrium temperatures
$T_h'$ and $T_c'$ for respectively $\R$ and $\S$ via $\rho\propto
e^{-\beta_h' H_\R}$ and $\sigma\propto e^{-\beta_c' H_\S}$. Thus,
$\beta_h'=\frac{1}{\bar{\eps}}\ln\frac{1}{\v}$ and
$\beta_c'=\frac{1}{\bar{\mu}}\ln\frac{1}{\u}$, where
$\v=e^{-\beta_c\bar{\mu}}$ and $\u=e^{-\beta_h\bar{\eps}}$; see
(\ref{burundi}).  This implies $T_cT_h=T'_cT'_h$.  As expected, the
refrigeration condition $\v>\u$, see (\ref{10}, \ref{14}), is equivalent
to $T_c'<T_c<T_h<T_h'$, i.e., the cold system gets colder, while the hot
system gets hotter. Note that the existence of temperatures $T'_c$ and $T'_h$ was not imposed, they emerged out
of optimization. 

We eventually focus on two important limits: quasi-equilibrium regime $\theta\to
1$, and the macroscopic regime $\ln n\gg 1$. 

{\it In the quasi-equilibrium regime} 
\BEA
\label{15}
\chi(a)|_{\theta=1}
= T_c \,\theta (n-1)\,
{[1+(n-1)a]^{-2}}\, \ln^2{a},
\EEA
maximizes for $\u=\v=a$, where $a$ is found from $\partial_a \chi(a)|_{\theta=1}=0$:
\BEA
\label{16} [{(n-1)a-1}]\, \ln a=2 [{(n-1)a+1}]
\EEA
We now work out the optimal $\u$ and $\v$ for $\theta\to 1$. It can be seen from (\ref{14})
that the proper expansion parameter for $\theta\to 1$ is $x\equiv \sqrt{1-\theta}$.
We represent $\u$ and $\v$ as
\BEA
\u=a+{\ssum}_{k=1}a_kx^k,~~~\v=a+{\ssum}_{k=1}(a_k+b_{k-1})x^k.\nonumber
\EEA
Substituting these expressions into $\partial_u \chi=0$ and $\partial_v
\chi=0$ and expanding these over $x$ we note that $a_k$ and $b_k$ are determined by 
equating the ${\cal O}(x^k)$ terms:
\BEA
b_0=a\ln\frac{1}{a}, \quad a_1=-\frac{a}{2}\ln\frac{1}{a}, \quad b_1=-\frac{a}{48}\ln\frac{1}{a}[24+\ln^2 a].\nonumber
\EEA
This implies for the efficiency at $\theta\to 1$ ($x=\sqrt{1-\theta}$)
\begin{gather}
\zeta= \frac{1}{x}-1+\frac{\ln^2 a}{48}-\frac{[48+\ln^2a]\ln^2 a}{1536}\,x
+{\cal O}(x^2).
\label{capo}
\end{gather}
Note that the expansion (\ref{capo}) does not apply for $
n\to\infty$, since $\ln a$ diverges in this limit; see (\ref{16}).

Eq.~(\ref{capo}) suggests that $\frac{1}{\sqrt{1-\theta}}-1$ is a lower
bound for the efficiency at the maximal $\chi$.  This is numerically
checked to be the case for all $0<\theta<1$ and all $n$; see also
Fig.~\ref{f2}.  Recalling (\ref{peshavar}) and our discussion after (\ref{14}), we can
interpret the lower bound for the efficiency as a lower bound on the
intermediate temperature $T'_c$ of $\S$: 
$\frac{1+\sqrt{1-\theta}}{\theta}<\frac{T'_c}{T_c}<1$, i.e., $T_c'$ cannot be too low.

{\it The macroscopic regime} of a $n$-level quantum system means $\ln (n-1)\gg 1$, since
for $N\gg 1$ weakly coupled particles the number of energy levels scales
as $e^N$. Now $\u$ and
$\v$ in (\ref{14}) are sought via the following asymptotic expansions ($m\equiv n-1$)
\BEA
\label{hawk}
\u={\ssum}_{k=1} {\rho_k}{[m\ln m]^{-k}}, ~~
\v={\ssum}_{k=1}\omega_k m^{-k}\ln^k m ,
\EEA
where $\rho_k$ and $\omega_k$ are found from substituting (\ref{hawk})
into $\partial_u \chi=0$ and $\partial_v
\chi=0$ and using $\ln (n-1)\gg 1$. In the first order we get 
$\rho_1=\frac{1}{1-\theta}$, $\omega_1=\frac{1-\theta}{2-\theta}$, which leads to
\begin{gather}
\label{tacit}
\zeta=\frac{\theta}{1-\theta}-\frac{2\theta}{(1-\theta)^2}\frac{\ln[\ln m]}{\ln m}+{\cal O}\left[ \frac{1}{\ln^2 m} \right],\\
\label{tacit2}
\frac{Q_c}{T_c}=\ln m-\frac{3-\theta}{1-\theta}-\ln\left[ \frac{1-\theta}{2-\theta}\, \ln m  \right]
+{\cal O}\left[ \frac{1}{\ln m} \right],  \nonumber
\end{gather}
It is seen that in the macroscopic limit the efficiency converges to the
Carnot value, while the transferred heat $Q_c$ is (in the leading order)
a product of the colder temperature $T_c$ and the "number of particles"
$\ln (n-1)$. Note that the obtained attainability of the Carnot bound is
related to a finite power and a finite $Q_c$. We see
that the macroscopic limit does not commute with the equilibrium
limit, since the corrections in (\ref{tacit}) diverge for $\theta\to 1$.

{\it Classical limit.} A
maximization of $\chi=Q_c\zeta$ can be carried out imposing {\it equidistant} spectra
$\eps_n=n\eps$ and $\mu_n=n\mu$ for $\R$ and $\S$. We find that
the optimal ${\cal U}$ again corresponds to
SWAP operation. Thus, for $\chi=\chi(\u,\v)$ we obtain
\BEA
\label{144}
\chi= 
\frac{T_c\theta \ln^2\frac{1}{\v}}
{\ln\frac{1}{\u}-\theta\ln\frac{1}{\v}}\left[
\frac{\v-\u}{(1-\v)(1-\u)}-\frac{n(\v^n-\u^n)}{(1-\v^n)(1-\u^n)}
\right],\nonumber
\EEA
where $\v=e^{-\beta_c\bar{\eps}}$ and $\u=e^{-\beta_h\bar{\mu}}$ are
found from maximizing $\chi$. The efficiency $\zeta$ is still given by (\ref{11}). In the limit $n\gg 1$ we get from
(\ref{144}): $\bar{u}\to 1$, $\bar{v}\to 1$ and
$\frac{n(\v^n-\u^n)}{(1-\v^n)(1-\u^n)}\to 0$.  Both $\chi$ and $\zeta$
depend on one parameter $\phi\equiv\frac{1-\bar{u}}{1-\bar{v}}$, whose optimal value
is $\phi=1+\sqrt{1-\theta}$.  We get in this limit:
$\chi=\frac{T_c\theta}{(1+\sqrt{1-\theta})^2}$ and
$\zeta=\frac{1}{\sqrt{1-\theta}}-1$.  Thus for a large number of
equidistant energy levels (macro-limit) the optimal regime now implies homogeneity
($\bar{\eps}\to 0$, $\bar{\mu}\to 0$), which is an indication of the classical
limit: under this conditional optimalization the efficiency $\zeta$ is
exactly equal to the [unconditional] lower limit
$\zeta_{\rm CA}=\frac{1}{\sqrt{1-\theta}}-1$.

{\it In conclusion}, we have studied a model of a refrigerator aiming to
understand its optimal performance at a finite cooling power; see
Fig.~\ref{f1}.  The structure of the model is such that it can be
optimized over almost all its parameters; additional constraints can and
have been considered, though. We have confirmed an incompatibility
between optimizing the heat $Q_c$ transferred from the cold bath $T_c$
and efficiency $\zeta$: Maximizing one nullifies the other.  A similar
effect for a different model of quantum refrigerator has been reported
in \cite{segal}. 

To get a balance between $Q_c$ and $\zeta$ we have thus chosen to optimize
their product $\zeta Q_c$. This leads to a {\it lower} bound $\zeta_{\rm
CA}=\frac{1}{\sqrt{1-\theta}}-1$ ($\theta\equiv \frac{T_c}{T_h}$) for
the efficiency in addition to the upper Carnot bound $\zeta_{\rm
C}=\frac{1}{1-\theta}-1$. The Carnot upper bound is
reached (at a finite power and finite $Q_c$!) in the macroscopic
(many-level) limit of the model. To our knowledge such an effect has
never been seen so far for refrigerator models. For the optimal refrigerator
the transferred heat $Q_c$ behaves as $Q_c\propto T_c$ for $T_c\to 0$;
see (\ref{10}, \ref{14}, \ref{tacit}). This is in agreement with the
optimal low-temperature behaviour of $Q_c$ from the viewpoint of the
third law \cite{rezek}. 
The lower bound $\zeta_{\rm
CA}$ is reached in the equilibrium
limit $T_c\to T_h$. Constraining both systems
to have homogeneous (classical) spectra, $\zeta_{\rm CA}$
is reached as an upper bound. This is just like 
within finite-time thermodynamics (FTT), when maximizing the product of the cooling-power and
efficiency \cite{yan_chen}, or the ratio of the
efficiency and the cycle time \cite{velasco}. 
In this sense $\zeta_{\rm
CA}$ seems to be universal. It may play the same role as the
Curzon-Ahlborn efficiency for heat engines $\eta_{\rm CA}$, which,
again, is an upper bound within FTT \cite{ca,broek}, but appears as a
lower bound for the engine models studied in
\cite{armen,tu,izumida_okuda}. Other opinions on the Curzon-Ahlborn
efficiency for refrigerators are given in \cite{jimenez,unified}. 

This work has been supported by Volkswagenstiftung.

\end{document}